# How to Avoid Reidentification with Proper Anonymization

**Authors:** David Sánchez*, Sergio Martínez, Josep Domingo-Ferrer

**Affiliations:**

UNESCO Chair in Data Privacy, Department of Computer Engineering and Mathematics, Universitat Rovira i Virgili (URV), Av. Països Catalans, 26, E-43007, Tarragona, Catalonia.

*Correspondence to: E-mail: david.sanchez@urv.cat.

**Abstract**: De Montjoye et al. (*1*) claim that most individuals can be reidentified from a de-identified transaction database, and that anonymization mechanisms are not effective against reidentification. We demonstrate that these claims are due to a misconception of the reidentification attack, a poor anonymization and a disregard of 40 years of anonymization literature. We also show how proper anonymization can be performed.

**One Sentence Summary:** We demonstrate that previous reidentifications claimed for transaction data were due to poor anonymization, and we show how to properly anonymize data.

**Main Text:**

De Montjoye et al. (*1*) (dM from now) concluded that, for most customers in a de-identified credit card transaction database, knowing the spatiotemporal features of four purchases by the customer was enough to reidentify her. Reidentification was measured according to "unicity" (*2*) (a neologism for the plain old uniqueness notion) which, given a number of personal features assumed known to an attacker, counts the number of individuals for whom these features are unique.

First, dM's reidentification figures are probably overestimated, because their database of 1.1 million customers seems only a fraction of the population of an undisclosed country (presumably, several millions). Unfortunately, dM did not make their dataset public, which prevents reproducing their results. As highlighted by Barth-Jones et al. (*3*), with a non-exhaustive sample, an individual's sample uniqueness/unicity does not imply population uniqueness and, hence, does not allow *unequivocal* reidentification; assuming otherwise clearly overestimates the reidentification risk. Moreover, not even population uniqueness automatically yields reidentification: the attacker still needs to link the records with unique features to external *identified* data sources (e.g., electoral rolls).

To reduce the high "unicity" in their database, dM implemented some [un-referenced] "anonymization" strategies to coarsen data (such as clustering locations) that fell short of sufficiently reducing "unicity". From this, dM drew bold conclusions about the ineffectiveness of anonymization methods and highlighted the need for *"more research in computational privacy"*.

We must recall that reidentification risk in data releases has been treated in the *statistical disclosure control* (*4,5*) and *privacy-preserving data publishing* (*6*) literatures for nearly four decades. As a result, a broad choice of anonymization methods exists, which dM systematically overlook. These usually suppress personal identifiers (such as passport numbers) and mask *quasi-identifiers* (QIs). The latter are those attributes (such as zipcode, job or birthdate) each of which does not uniquely identify the subject, but whose combination may. Since QIs may be present in public non-confidential databases (like electoral rolls) together with some identifiers (like passport number), it is crucial to mask them to avoid reidentification. It is easy to see that reidentification via QIs (studied at least since 1988 (*7*) and popularized by the *k*-anonymity model (*8*)) is equivalent to the "unicity" idea re-discovered by dM in 2013 (*2*); that is, a subject whose QI values are unique in a dataset risks being reidentified.

Data coarsening is indeed a method often used in anonymization to mask QIs (*8*). However, dM concluded that their coarsening-based anonymization was ineffective. This is unsurprising because they coarsened attributes independently and used value ranges fixed *ex ante*, which is naïve and inappropriate for at least two reasons: (i) to offer true anonymity guarantees, coarsening should be based on the actual distribution of the dataset (i.e., a fixed range may contain a single value among those in the dataset); (ii) independently coarsening each QI attribute cannot ensure unique QI value combinations disappear (coarsening must consider all QIs together).

To illustrate the effectiveness of *sound* anonymization, the simple and well-known *k*-anonymity notion is enough. In a *k*-anonymous dataset, records should not include strict identifiers, and each record should be indistinguishable from, at least, *k*-1 other ones regarding QI values. Thus, the probability of reidentification of *any* individual is $1/k$. Hence, for $k>1$, this probability is less than 1 for *all* records, thereby ensuring *zero unequivocal* reidentifications. Moreover, by tuning *k*, we can also tune the level of exposure of individuals.

dM's data being withheld, we were forced to look for a dataset with similar structure and "unicity"/reidentification risk properties to show the effectiveness of *k*-anonymity. We chose a synthetically generated version SPD of a publicly available patient discharge dataset that includes the nearly 4 million patients admitted in 2009 to Californian hospitals (see details in the supplementary materials (*9*)). This dataset includes a set of spatiotemporal features of the patients and, unlike dM's dataset, it covers the *whole* population of 2009 Californian patients; hence, uniqueness in this dataset *truly quantifies* the population reidentification risk (see (*9*)). As shown in Figure 1 the high risk reached for the SPD dataset when the attacker knows all the patient's features (75%) is coherent with the high (even though overestimated) "unicities" reported by dM.

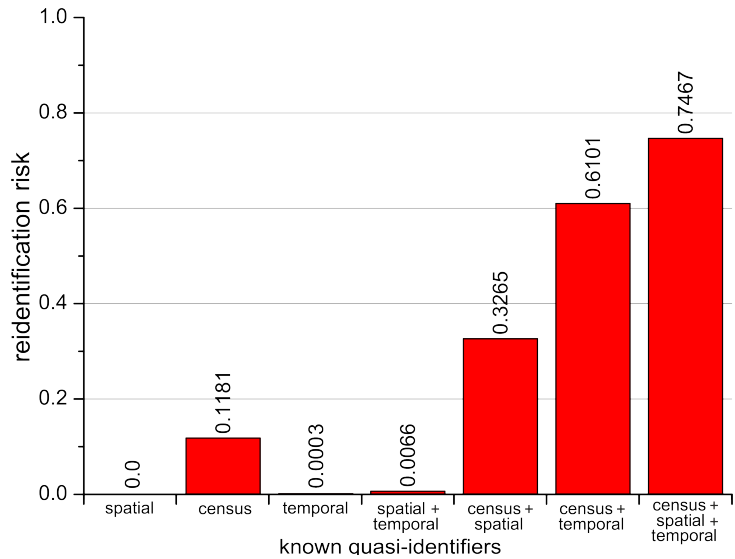

**Fig. 1. Reidentification risk in the SPD dataset depending on the attributes known by the attacker**.

We enforced *k*-anonymity by grouping records with *similar* QIs (census+spatiotemporal features) in clusters of *k* or more, and generalizing/coarsening their QI values to their common range (*9*). Figure 2 compares the risk of unequivocal reidentification and correct random reidentification of *k*-anonymity vs. a naïve coarsening similar to dM's, with "fixed" intervals covering 1/32, 1/16 and 1/8 of the domain ranges of the attributes (see (*9*)). Unlike naïve coarsening, *k*-anonymity yields *zero* unequivocal reidentifications and a rate 1/*k* of correct random reidentifications when the attacker knows all QIs.

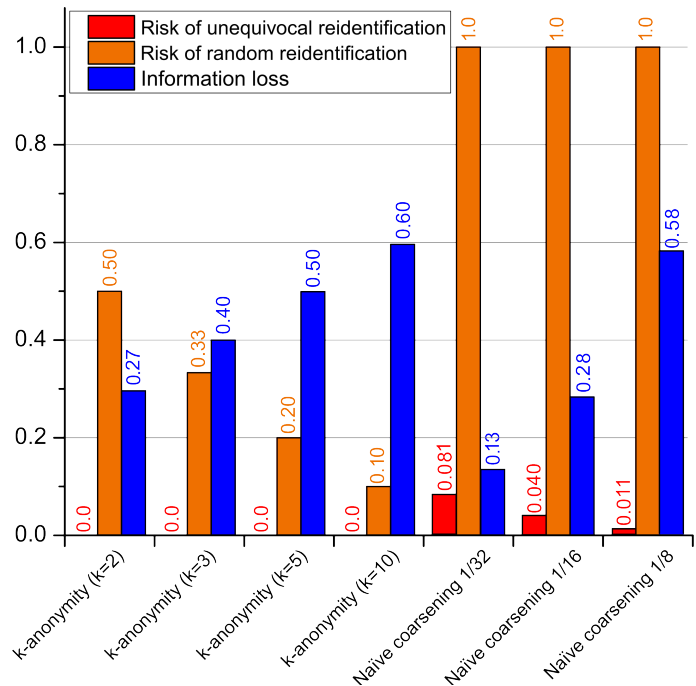

**Fig. 2. Reidentification risk and information loss for *k*-anonymity and naïve coarsening.**

Furthermore, anonymized data should also retain analytical utility, which ultimately justifies data publishing. With *k*-anonymity, data utility is retained by grouping *similar* records together and by masking only those that do not fulfill the privacy criterion (dM's naïve coarsening fails to do either). Moreover, the trade-off between privacy and utility can be balanced by adjusting *k*. To illustrate, we have measured the *information loss* incurred by masking as the average distance between the SPD dataset and its anonymized versions (*9*). Figure 2 shows that 2-anonymity not only yields less reidentifications, but also less information loss than the safest naïve coarsening.

In addition to *k*-anonymity, there is much more in the anonymization literature. Specifically, extensions of *k*-anonymity (e.g., *t*-closeness (*10*)) also address *attribute disclosure*, which occurs if the values of the confidential attributes within a group of records sharing all QI values are too close. In (*9*) we report how *t*-closeness mitigates attribute disclosure by using the algorithm we proposed in (*11*). Moreover, the current research agenda includes more challenging scenarios, like *big data* anonymization (in which scalability and linkability preservation are crucial) (*12, 13*), streaming data anonymization (*14*) and local or co-utile collaborative anonymization by the data subjects themselves (*15*).

In conclusion, data owners and subjects can be reassured that sound anonymization methodologies exist to produce useful anonymized data that can be safely shared for research.

**Acknowledgments:** Supplementary materials to this study are available at http://arxiv.org/abs/1511.05957. They detail the structure and synthetic generation of the SPD dataset, describe the risk assessment and anonymization algorithms we used to obtain the reported results, and provide extended results and discussions. The SPD dataset, with the synthetic quasi-identifiers, and its *k*-anonymous, k-anonymous & *t*-close, and coarsened versions can be found at http://crises-deim.urv.cat/opendata/SPD_Science.zip. The source code of the algorithms detailed in the supplementary materials is also available together with some usage examples. These materials allow reproducing all results reported here and in the supplementary materials.

Thanks go to Jane Bambauer, Ann Cavoukian, Khaled El Emam, Krish Muralidhar and Vicenç Torra for useful reviews and discussions.

The following funding sources and grants are gratefully acknowledged: European Commission (H2020-644024 "CLARUS"), Spanish Government (TIN2012-32757 and TIN2014-57364-C2-1-R), Government of Catalonia (2014 SGR 537 and ICREA-Acadèmia award to J. Domingo-Ferrer), and Templeton World Charity Foundation (TWCF0095/AB60). The opinions expressed in this paper are the authors' own and do not necessarily reflect the views of any funder or UNESCO.